\documentclass[prl,superscriptaddress,twocolumn]{revtex4-1}
\pdfoutput=1
\usepackage[colorlinks=true,urlcolor=blue]{hyperref}
\usepackage{amsfonts}
\usepackage{graphicx}
\usepackage{mathrsfs}
\usepackage{amsmath}
\usepackage{upgreek}
\usepackage{xcolor}
\usepackage{bm}
\usepackage{titlesec}

\usepackage{color}
\definecolor{red}{rgb}{0.75,0,0}
\definecolor{blue}{rgb}{0,0,0.75}
\definecolor{green}{rgb}{0,0.5,0}

\titlespacing*{\section}
{0pt}{0.5cm}{0.25cm}
\titlespacing*{\subsection}
{0pt}{0.5cm}{0.25cm}

\begin{document}

\title{Microbial Active Matter: A Topological Framework}

\author{Anupam Sengupta}
\email{anupam.sengupta@uni.lu}
\affiliation{Physics of Living Matter Group, Department of Physics and Materials Science, \\ University of Luxembourg, 162 A, Avenue de la Faïencerie, \\ L-1511 Luxembourg City, Luxembourg}
\date{\today}

\begin{abstract}
\vspace*{0.5cm}
Topology transcends boundaries that conventionally delineate physical, biological and engineering sciences. Our ability to mathematically describe topology, combined with our access to precision tracking and manipulation approaches, has triggered a fresh appreciation of topological ramifications, specifically in mediating key functions in biological systems spanning orders of magnitude in length and time scales. Microbial ecosystems, a frequently encountered example of living matter, offer a rich test bed where the role of topological defects and their mechanics can be explored in the context of microbial composition, structure and functions. Emergent processes, triggered by anisotropy and activity characteristic of such structured, out-of-equilibrium systems, underpin fundamental properties in microbial systems. An inevitable consequence of anisotropy is the long-range orientational (or positional) correlations, which give rise to topological defects nucleating due to spontaneous symmetry breaking. The scene stealer of this emerging cross-disciplinary field is the topological defects – singularities embedded within the material field that elicit novel, if not unexpected, dynamics that are at the heart of active processes underpinning soft and living matter systems. In this short review, I have put together a summary of the key recent advances that highlight the interface of liquid crystal physics and the physical ecology of microbes; and combined it with original data from experiments on sessile species as a case to demonstrate how this interface offers a biophysical framework that could help to decode and harness active microbial processes in \textit{true} ecological settings. Topology and its functional manifestations – a crucial and well-timed topic – offer a rich opportunity for both experimentalists and theoreticians willing to take up an exciting journey across scales and disciplines.  \bigskip
\end{abstract}

\maketitle

\section{\label{sec:intro} Introduction}

\begin{figure}[t]
\centering
\includegraphics[width=\columnwidth]{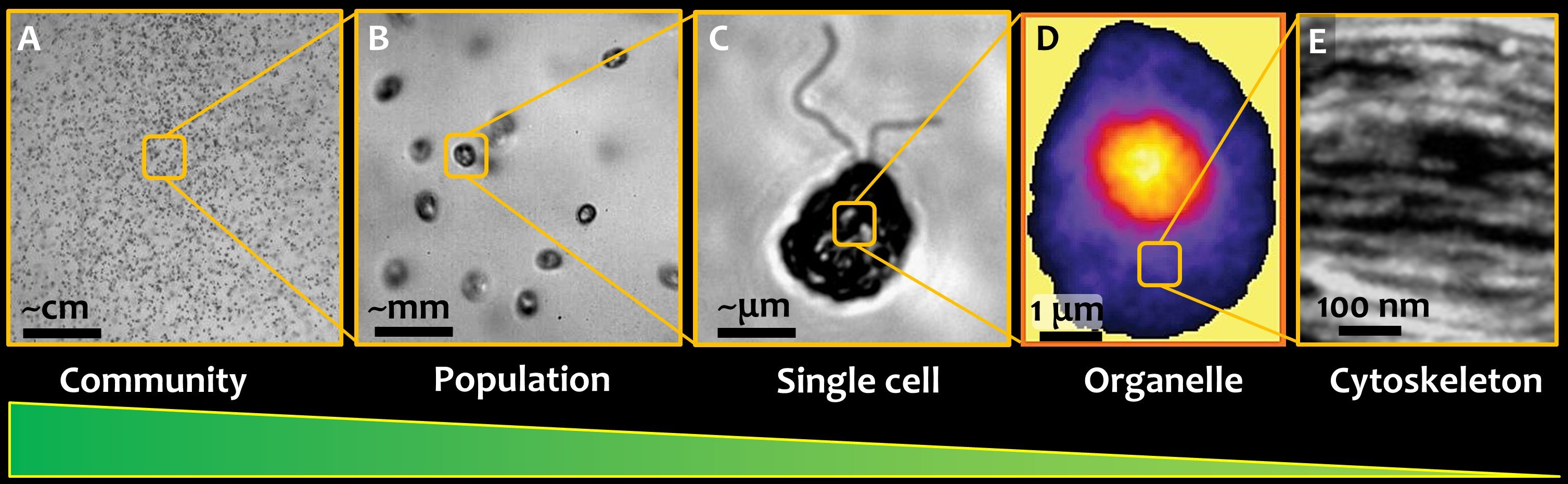}
\caption{\label{fig:exp} {\bf Microbial Active Matter.} Microbial active matter spans multiple scales: from communities comprising populations of multiple species (A), to microbial populations of single species (B), down to individual cells (C). Out-of-equilibrium physics underpins biophysical processes at sub-cellular scales, regulating phase separation, organelle compartmentalization and microbial shape-shifting (D) and directional molecular transport by the cytoskeletal elements (E). The scale bar in green represents the decreasing system length scales on moving form the micorbial community to the cytoskeletal elements. Panel D is adapted from the author’s work \cite{Sengupta:2017}, and panel E from \cite{Jin:2008}, with permissions from the Nature Publishing Group and John Wiley and Sons respectively.} 
\end{figure}

Microbes mediate and dictate a broad range of processes in ecology, medicine and industry. The urgent need for devising better antibiotics, the development of bioremediation approaches for anthropogenic disasters such as oil spills, application of microbes towards sustainable ecosystems, and the need to coherently assess how microbes govern the dynamics of soil, plant, marine and human ecosystems – all require an articulate understanding of the vital functions that microbes carry out. Microbial activity spans multiple scales: from community dynamics playing over millimeter to meter scales, down to sub-cellular organelles with characteristic lengths of hundreds of nanometers Fig.\ref{fig:exp}. A significant proportion of these microbes – from prokaryotic bacteria making up different biotopes, to eukaryotic phytoplankton in flow – occupy highly dynamic natural habitats, where a combination of periodic and stochastic variations in their micro-environments shape the species fitness, succession, and selection\cite{Shade:2012, Palme:2012}. Since Pasteur formalized the nexus between microbiology and materials (in this case, food materials)\cite{Pasteur:1872}, the scientific and industrial pursuit of biotechnology at the interface of microbes and materials has continued unhindered. This lasting advancement was realized in part due to the discovery of diverse microbial taxonomies within different contexts\cite{Hungate:1979,Adler:2018,Dewit:2006}, and elucidation of the intricate community structures therein, also known as the microbiota, or more commonly, microbiome\cite{Peterson:2009,Turnbaugh:2007}. Alongside, a close understanding of the biophysical attributes of the environment has enabled valuable insights into microbial behavior and physiology in a dynamic environment. At the scale of a microorganism, the local micro-environment can be generalized as a spatially structured complex soft material, with internal energies spanning equilibrium thermal energies ($k_{B}T$, the product of the Boltzmann constant, $k_{B}$, and the absolute temperature, $T$),\cite{Mezzenga:2005,Jansson:2020,Kim:2019} to out-of-equilibrium active environments\cite{Ramanan:2016,Sommer:2017,Sengupta:2017}. Ranging from 1 nm to 100 μm – five orders of length scale – micro-structural complexity coexists with microbial diversity in vast majority of natural and nature-inspired microbial ecosystems. Key microorganisms, or the core microbiota, from a range of applied microbial settings, have yielded plethora of information on optimal physiology and fitness, relevant from a fundamental microbial perspective\cite{Tshikantwa:2018}. Together with the rapid progress in sequencing and omics tools, this has led to a systematic and high throughput analysis of microbial metabolism and response pathways\cite{Lederberg:2001}. Yet microbiology and microscale physics have rarely been considered as an ensemble – a single composite biophysical system – that underpins the natural and synthetic microbial processes. Bulk of the existing studies – both experimental and modeling – have considered one or the other, and thus, relatively little is known about the active biophysics that govern the microbiome dynamics in general, and the microbe-environment interactions in particular. Specifically, by analyzing microbial ecosystems through the lens of active matter physics, two distinct uncharted biophysical themes emerge: (1) Activity and emergence in microbial consortia: how emergent properties are triggered (or hindered) in communities of multiple players (species) with distinct biophysical traits; and (2) Microbial behaviour and physiology in relation to the dynamic micro-environments they are part of. In other words, can we harness environmental dynamics to tune microbial activity and emergent properties? Both the themes, interfacing microbial ecology and active matter physics, have this far went unexplored, despite their relevance and potential impact. The holy grail will be to develop a mechanistic framework that could decouple the two scenarios and reveal the relative influence of the community composition and structure, vis-a-vis the environmental dynamics and attributes. In an integrative approach, species in a consortium could be considered a part of the microbial environment itself. Nonetheless, an unambiguous understanding of each species in a microbial community, and their relation to the micro-environment, will be crucial in assessing their contribution to the environmental variables. 

In a biophysical context, microbes can be generalized as microscale biological active matter that expends energy to perform tasks and processes information to execute physiological functions, ultimately enabling them to maintain biological fitness. Microbes have thus been considered as model systems, based on which theories for active matter systems have been developed\cite{Bechinger:2016,Fodor:2018,Marchetti:2013,Cates:2015,Klamser:2018}. Broadly classified under prokaryotes (unicellular organisms without membrane-bound nucleus) and eukaryotes (uni- or multicellular organisms possessing membrane-bound nucleus), microbes can be planktonic (motile) or sessile (non-motile), inhabiting different ecosystems. Motility imparts cells the ability to actively propel, aided by plethora of propulsion mechanisms\cite{Lauga:2016}. On the other hand, non-motile microbes could be surface attached, or rely on passive mechanisms for locomotion. Over the recent years, there has been a growing interest to understand the dynamics of microbial systems with higher complexities: coexisting motile and non-motile species\cite{Coelho:2020,Patteson:2018,Xu:2019}, microbes in complex fluids\cite{Riley:2014,Zottle:2019,Makarchuk:2019}, and active response and feedback between microbes and their micro-environments\cite{Sommer:2017,Sengupta:2017,Bechinger:2016,Kranz:2016}. Despite the unprecedented progress over the last decade, the field of microbial biophysics faces conceptual challenges on the way, specifically in linking the physics of active matter to the biology of microbial ecosystems in natural or nature-inspired ecosystems. A close scrutiny would reveal that microbes, as they exist today in their natural habitats, have emerged from eons evolution, guided by the interplay of physics and genetics\cite{Sackmann:2013}. Thus, for a consequential application of the theory in true biological settings, an appreciation of of the role of the environmental variations and the underlying molecular pathways, in addition to the material and mechanical attributes of the cells, will be vital. Engaging biomolecular approaches in tandem with microbial biophysics will trigger an iterative feedback where the knowledge of biological pathways will inform new and update existing theory, and vice versa; ultimately enabling predictive approaches for microbial biophysics\cite{Scheffer:2012,Goldford:2018}. Furthermore, bridging theory with the biophysical experiments has been hindered by the multiplicity of microbial traits and processes that act simultaneously, affecting both the consortia members and the micro-environment\cite{Goldford:2018}. Our ability to move from experiment-specific theory to general principles of microbial biophysics could garner much wider attention from both aisles of the scientific community, ultimately offering an integrative framework for microbial ecology. These challenges have offered untapped opportunities for the active matter community, which could allow existing theories to be validated and iteratively updated to capture true biological systems. The objective of this article is three-fold: (1) to summarize key recent works on microbial active matter, with a focus on the role of geometric and topological features of the microbes and their surroundings; (2) present selected original results that capture the topological facets in microbial systems, in particular sessile bacteria; and (3) conclude with a perspective on the topological framework, and its promise in future studies on microbial systems.

\section{\label{sec:perspective}Microbial ecology: a topological perspective}

Microbes occupy every part of our biosphere, often as a part of complex community structures, interacting closely with dynamic micro-environments via physico-chemical cues. The ability of microbes to respond and adjust to environmental changes spanning vastly different scales (individual to community scales; from generational to evolutionary time scales), is a conundrum that has long intrigued biologists and physicists alike. Consequently, understanding how microbes interface, exchange and communicate with their local surroundings is central to the grand quest for a theory of microbial ecology. On the timeline of microbial ecology, it is rather recent that cellular and sub-cellular biophysics has started to emerge as a key player, propelling our understanding of microbial lifestyles and strategies under biotic and abiotic variations in their environment. Recent advances in molecular and imaging techniques have started to uncover the functional role of active biophysics in microbial ecosystems, specifically in the context of topological defects\cite{Hartmann:2019,Kawaguchi:2017,Saw:2017}, yet we lack a mechanistic framework that could explain, generalize, and predict microbial fate under environmental fluctuations. Microbiology and microscale biophysics work in tandem in everyday ecological settings, though they have rarely been considered as an ensemble parameter for analysing microbial ecology. Of particular significance to microbe-environment interactions is phenotypic plasticity\cite{Miner:2005} – the ability of microbes to dynamically tune biophysical attributes, namely, morphology, cell size, motility or surface-association, without altering the genotype (i.e., the genetic makeup remains same). Variations in phenotypes, the composite of observable characteristics or traits in an organism, arise due to differential expressions of the genetic code due to interactions with the environment\cite{Kussell:2005,Maynard:2019,Haas:2019,Ackermann:2019}. Its fundamental role in establishing a link between organisms and their environment has been reported in all forms of life: from simple unicellular bacteria\cite{Ackermann:2019} and photosynthetic phytoplankton\cite{Sengupta:2017,Jin:2008} to highly organized multicellular eukaryotes\cite{Forsman:2015}. 

\begin{figure}[t]
\centering
\includegraphics[width=\columnwidth]{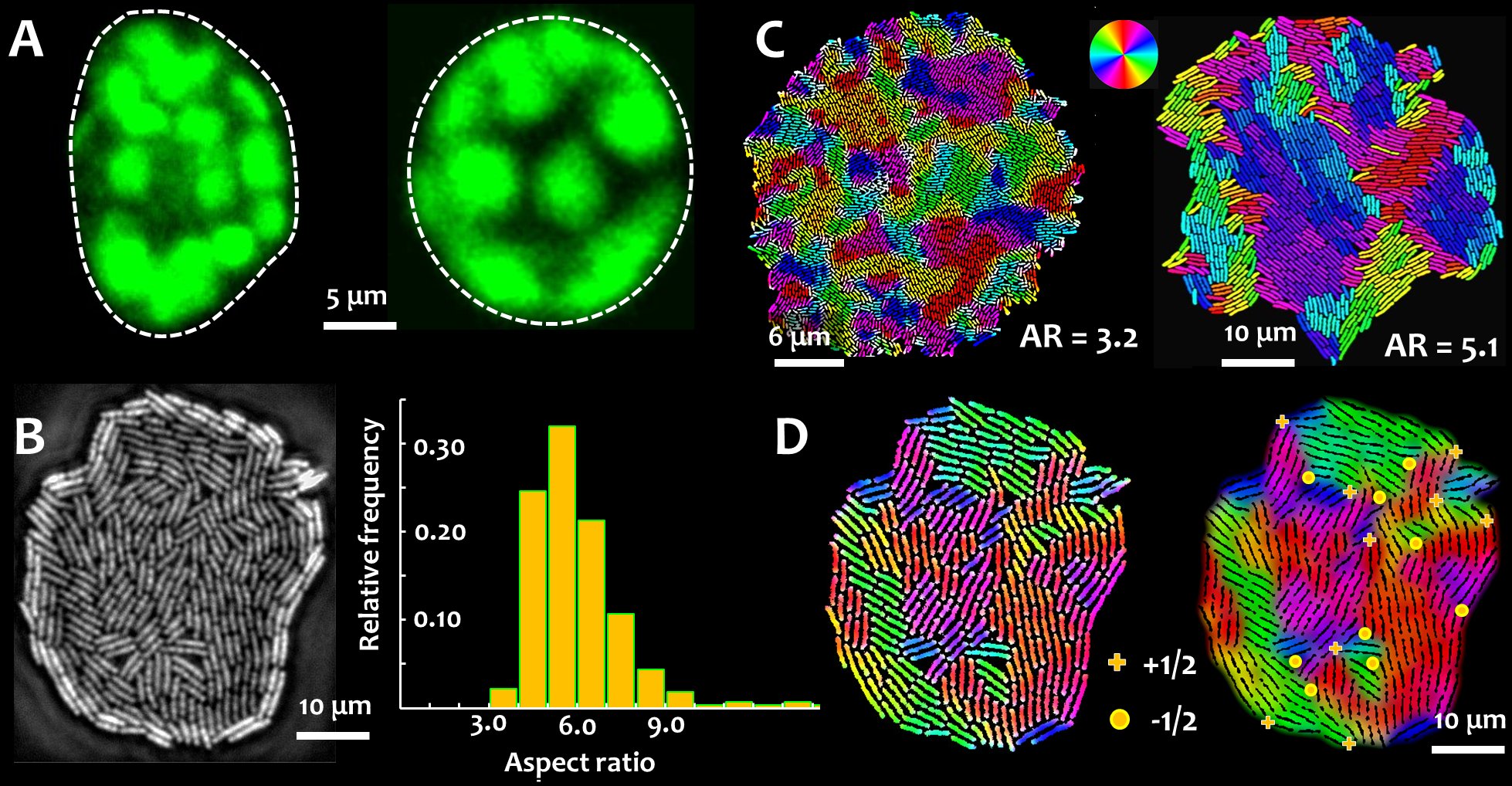}
\caption{\label{fig:snapshots} {\bf Microbial shape: a critical determinant of emergent properties.} Shape and its modulation are vital for microbial physiology and behaviour. (A) Phytoplankton can carry out rapid morphological transformation from asymmetric (left) to symmetric (right) shape as a response to hydrodynamic cues. Such transformations are mediated by the active reorganization of the cytoskeleton. (B) In an expanding colony of rod shaped bacteria, an intrinsic heterogeneity emerges in the aspect ratio of the cells. The growth conditions play a fundamental role in tuning the shape heterogeneity. (C) Cell aspect ratio determines the size of self-organized nematic microdomains in a growing bacterial colony. The cells are colour coded with respect to their local orientations, following the scheme shown by the colour wheel. (D) Emergent topological defects in an expanding bacterial colony. ($+$) sign indicates $+1/2$ integer defects and ($\bullet$) indicates $-1/2$ topological defects, which appear at the intersection of nematic microdomains. (Panel C is adapted from the author’s work \cite{You:2018} under the Creative Commons CCBY license).}
\end{figure}

At a functional level, phenotypic plasticity imparts individual cells, or populations, the capacity to cope with physiological requirements (for instance, necessitated by cellular age), or changes in the environmental conditions (e.g., response and adaptation). In the context of active matter physics, plasticity of phenotypes is analogous to tunable activity spanning different timescales – either at individual or collective levels. Figure \ref{fig:snapshots} presents the relative scales of organization in aquatic phytoplankton. An example of phenotypic plasticity – asymmetric and symmetric morphotypes – that emerge rapidly in motile phytoplankton exposed to turbulent hydrodynamic cues\cite{Sengupta:2017} is shown in Fig. \ref{fig:snapshots}A. The transition from an asymmetric to a symmetric cell shape depends on the properties of the external cue and the physiological state of the cells, which as depicted in Fig. \ref{fig:exp}E, is potentially mediated by the intracellular cytoskeletal matrix. The nature of the cytoskeletal element, in combination with the orientational order of the cytoskeletal network, determine the cell geometry, playing a fundamental role in sensing and transmission mechanical perturbations\cite{Fletcher:2010,Liu:2017}. Physiologically, cytoskeletal organization and its dynamics mediate crucial functions in marine microorganisms\cite{Durak:2017, Tyszka:2019,Gemmell:2016,Matt:2016}. Recent reports have suggested that cytoskeletal organization regulates biomineralisation\cite{Durak:2017}, and shapes mineralized\cite{Tyszka:2019} and labile\cite{Gemmell:2016,Matt:2016} forms. The secretion of biomineralised elements is tuned – in contrasting manners – by the disruption of the actin and microtubule networks. However, if and to which extent cytoskeletal order and emergent topological constraints, contribute to the transport processes involved, are yet to be explored. This could be particularly interesting in light of anisotropic diffusion in intra-cellular environments, known to facilitate the encounter and interaction of spindle-associated proteins\cite{Pawar:2014}. Linking cytoskeletal order and organization to cellular physiology and functions will enable an integrative understanding of microbial behaviour and lifestyles, while furthering our comprehension of multiscale complexities in nature for potential material science applications.

Phenotypic variations are commonplace among prokaryotes too, both at single cell and population scales. In absence of changes in their micro-environment, individual cells in a microbial population can exhibit intrinsic phenotypic heterogeneity\cite{You:2018,You:2019}. As shown in Fig.\ref{fig:snapshots}B, within an expanding colony of bacteria under steady conditions (nutrients and temperature), cells can exhibit intrinsic differences in the cell aspect ratio, i.e. the ratio of length to width of the cell. Additionally, phenotypic heterogeneities can emerge as a consequence of the ecological constraints, with different cell morphologies competing for limited resources. Recent results suggest that microbial populations comprising morphologically distinct shapes can undergo spatial structuring\cite{Smith:2017}, with potential ramifications on the cell lineage and fitness. Additional heterogeneities in phenotypes can co-exist, or co-emerge, including different cell structures (e.g., size and number of flagella), motility, surface attributes (e.g., adhesive properties) and growth rates. Taken together, phenotypic traits and variations therein, lead to a rich biophysical landscape where the interplay of microbial activity, geometry and local order trigger emergent behavior and functions. 

\begin{figure*}
\centering
\includegraphics[width=0.80\textwidth]{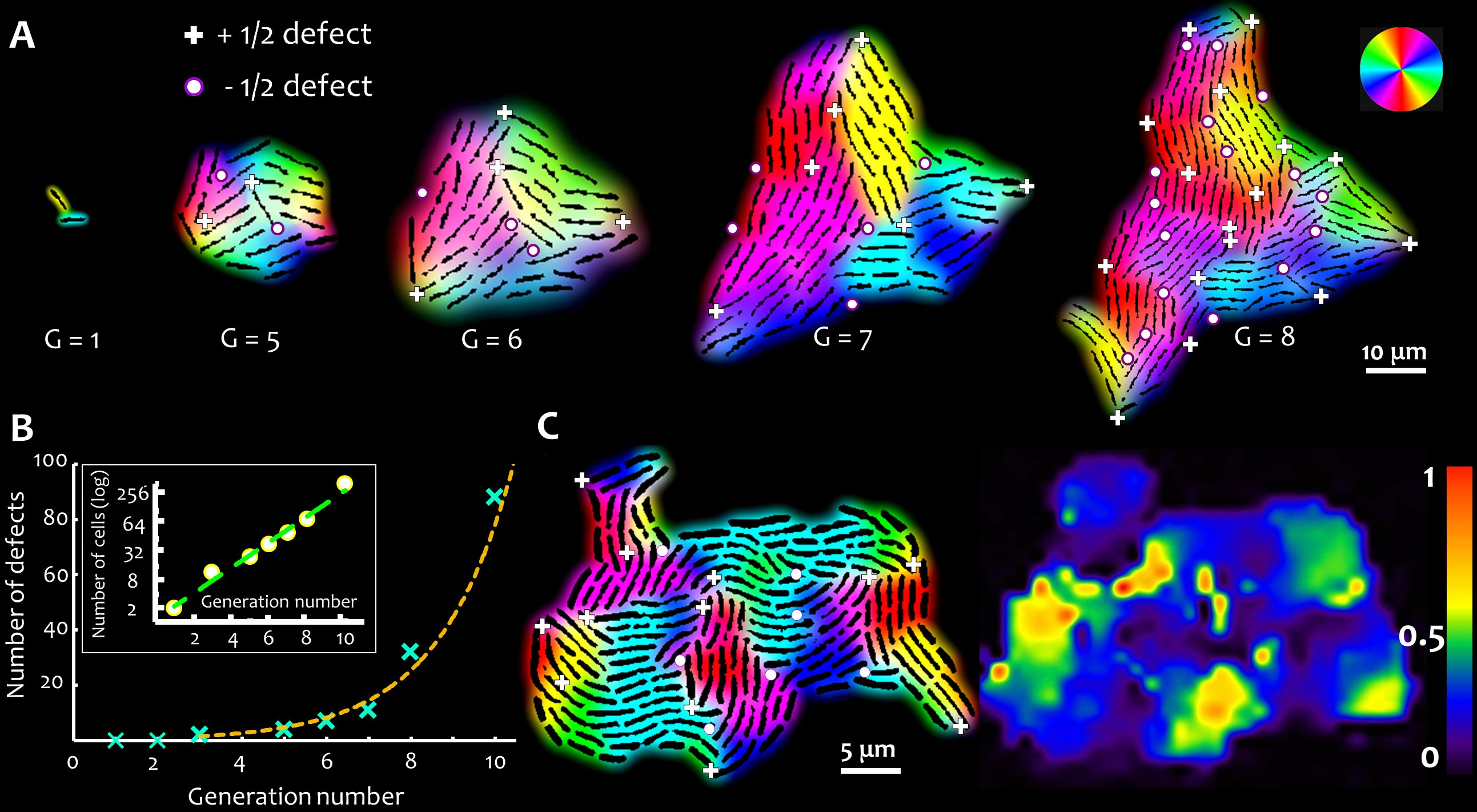}
\caption{\label{fig:stress} {\bf Topological defects in a growing bacterial colony.} Close packing of microbes with anisotropic shape trigger spontaneous formation of topological defects. (A) Time lapse snapshots of a growing bacterial colony of non-motile $E. coli$ strain (NCM 3722 delta-motA), where $G$ refers to the generation time (i.e., the number of cell doubling events). Here the cells divide every $\sim$45 minutes. As the colony expands, topological defects emerge spontaneously, either at the periphery or within the bulk of the colony. The defects nucleate at the intersection of 3 or more distinctly oriented microdomains, which are indicated here by ($+$) and ($\bullet$) signs, for $+1/2$ integer and $-1/2$ integer topological strengths respectively. (B) The total number of topological defects (sum of $+1/2$ and $-1/2$ integer defects) increases close to exponentially with time (represented here by the generation number, $G$), and near linearly to the corresponding cell number in the colony. The corresponding bacterial cell numbers are plotted on a log scale, in the inset. (C) As the cells keep dividing, the growth-induced mechanical stresses reorganize the topological defects within the colony, triggering active cell flows within the colony. The cell flows are quantified using a particle image velocimetry, revealing a patchy flow landscape with counter-rotating vortical regions. The relative strength of the active flows is indicated by the heat map scale bar. }
\end{figure*}

For surface-associated bacterial colonies, the generation and propagation of growth-induced active stress is determined by the activity (cell division rate) and the cell geometry (Fig.\ref{fig:snapshots}C). The emergence of microdomains is mediated by two competing forces: the steric forces between neighboring cells and the extensile stresses due to cell growth, which respectively tend to favor cell alignment and disrupt the local orientational order of the system\cite{You:2018}. The aspect ratio of cells in a colony determines the overall size (area) and spatial distribution of emergent nematic microdomains. Thus, for a given number of bacterial cells, an increase in the aspect ratio results in fewer number of nematic microdomains, each of which is however larger in size. The interplay of growth stresses and the steric interactions results in an exponential distribution of the domain areas, with a characteristic lengthscale proportional to the square root of the ratio between the orientational stiffness of the system and the magnitude of the extensile active stress. For a given size of the colony, the tradeoff between the size of the bacterial cells and that of the nematic microdomains determine the total number of topological defects. The defects nucleate at the intersection of three (or occasionally, four) distinctly oriented nematic microdomains, as shown in Fig.\ref{fig:snapshots}D, where the $+1/2$ and $-1/2$ topological defects are indicated by the ($+$) and ($\bullet$) signs respectively. The position and nature of the defects can be determined using a two-step analysis involving identification of the microdomain intersections (step 1), and then evaluating the angular rotation of bacterial cells around this intersection over a physical rotation of 2$\pi$ around the same point (step 2). Figure \ref{fig:stress}A and \ref{fig:stress}B respectively track the nucleation and the number of topological defects emerging in an expanding bacterial colony over multiple generations (indicated by $G$). Using time lapse imaging technique (imaging the colony at regular time intervals), the growth rate and dynamics of the defects can be studied over multiple generations. The total number of topological defects (sum of $+1/2$ and $-1/2$ integer defects) increases non-linearly over time, with a rate proportionate to the exponential growth of the colony (shown in the inset of Fig.\ref{fig:stress}B). 

The constellation of topological defects and their dynamics within colonies of different bacterial morphologies have been described as two and three dimensional active nematic systems\cite{You:2019,Doostmohammadi:2016,Arciprete:2018,Yaman:2019}. Theoretically, the shape of growing bacterial colonies was explained using continuum approach wherein cells were treated as active gel growing in an isotropic liquid\cite{Doostmohammadi:2016}. Friction, between cells and with the underlying substrate, was found to be a key determinant of the defect dynamics, which ultimately regulated the colony morphology. Growth of bacterial monolayers under soft agarose surfaces demonstrated that topological defects were created at a constant rate, with the motility of $+1/2$ defects biased towards the colony periphery\cite{Arciprete:2018}. More recently, studies on bacterial monolayers were extended to analyse multilayer morphologies, capturing the early developmental stages of bacterial biofilms\cite{You:2019,Yaman:2019,Beroz:2018}. Analytical modeling and numerical simulations have revealed that the transition from mono to multilayered morphology (in bacterial colonies of rod-shaped cells) is triggered by a competition between the growth-induced in-plane active stresses and vertical restoring forces due to the cell-substrate interactions\cite{You:2019}. Although the transition is localized and mechanically deterministic for small colony sizes, asynchronous cell division renders the process stochastic in larger colonies. In the limit of high cell numbers, the occurrence of the first division in the colony can be approximated as a Poisson process, the rate of which gives the order parameter of the transition, revealing the mixed deterministic-stochastic nature of the process. For bacterial colonies of chain-shaped cells, the multilayered structure emerged due to an interplay of mechanical stress accumulation and friction, resulting in buckling and edge instabilities\cite{Yaman:2019}. The buckling sites were characterized by nucleation of topological defects that initiated the formation of three-dimensional sporulation points. 

\section{\label{sec:flows}Active Microbial Flows: Dynamics of Topological Defects }

The spontaneous formation of nematic microdomains in an expanding bacterial colony nucleates topological defects in the local orientational field. In passive liquid crystals, the coupling between the hydrodynamic and nematic fields determines the dynamics of the topological defects and their influence on the viscoelastic properties of the fluid\cite{Sengupta:2014}. This often leads to exotic hydrodynamic ramifications: charge-dependent defect speeds, low Reynolds number cavitation phenomenon and coupling between singularities across disparate fields\cite{Toth:2014,Gimoi:2017,Stieger:2017}. Topological defects in active nematic systems differ fundamentally from their passive counterparts, on the following two fronts\cite{Giomi:2014}: i) defects in active nematic systems act as motile self-propelled particles with their motility (speed) proportional to the activity; and ii) defects in active nematic systems can nucleate continuously due to the local energy input. Consequently, the total number of topological defects (or defect pairs) keeps increasing with time. As defect tracking experiments have revealed\cite{Arciprete:2018,Doostmohammadi:2018}, $-1/2$ and $+1/2$ topological defects possess different intrinsic motilities: while $-1/2$ defects are observed to be nearly passive, advected with the expansion of the colony, the $+1/2$ defects have a sustained biased motility, the direction of which is determined by the extensile nature of the active stresses\cite{Doostmohammadi:2016,Giomi:2014}.

\begin{figure}[t]
\centering
\includegraphics[width=\columnwidth]{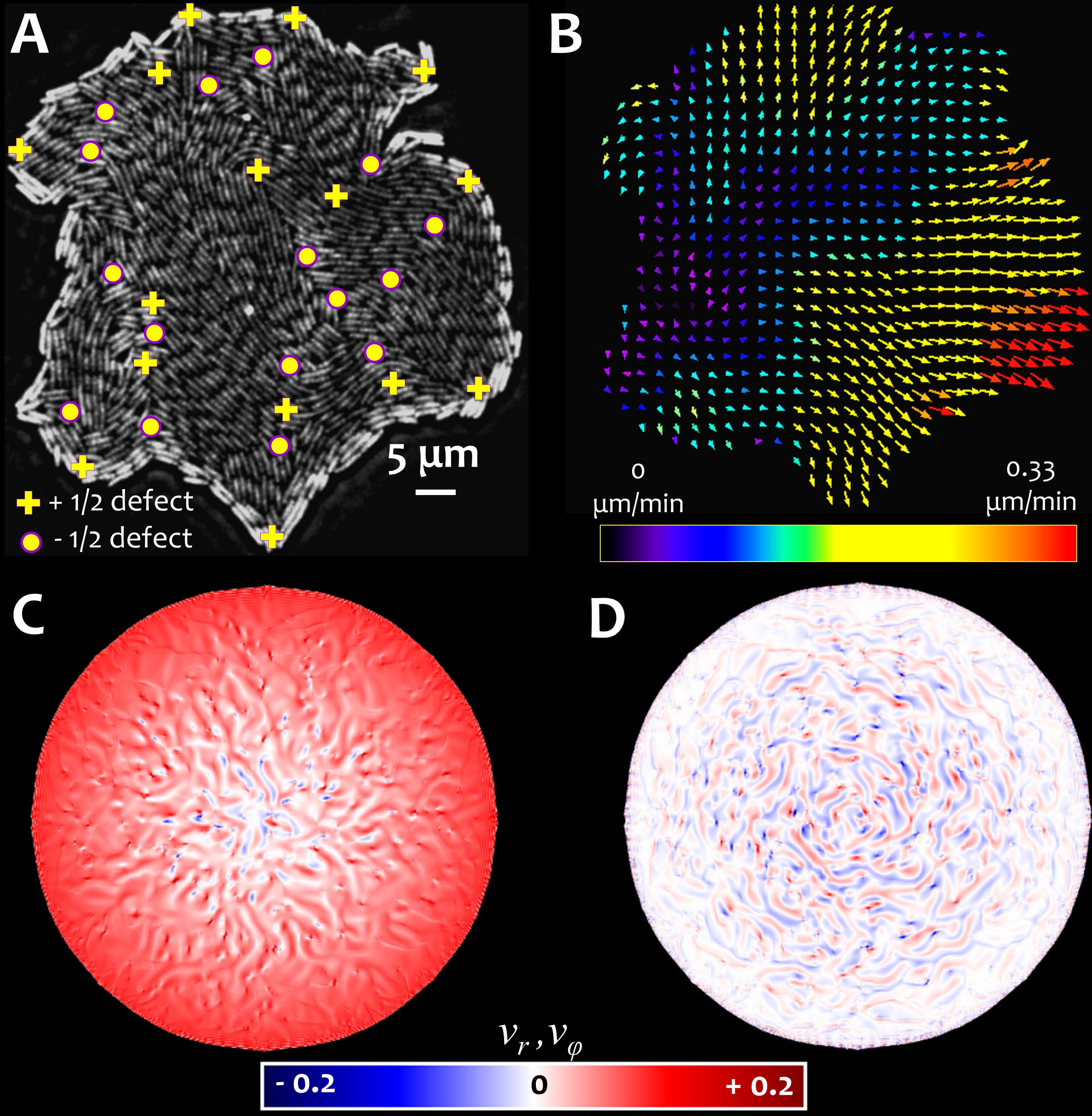}
\caption{\label{fig:statistics}{\bf Defect induced emergent flows in growing bacterial colonies.}(A) Micro-scale experiments capture a constellation of topological defects in a growing bacterial colony of non-motile $E. coli$ strain (NCM 3722 delta-motA). (B) The global active flow field, quantified using image velocimetry technique, where the arrow-heads indicate the direction of the emergent flow, colour coded with respect to the flow speed. Higher flow speeds are observed in regions with larger concentration of the topological defects. (C) Emergent hydrodynamic flows (radial and azimuthal components) simulated in an expanding bacterial colony. Panel C is adapted from author’s work\cite{You:2018} under the Creative Commons CCBY license. }
\end{figure}

As the cells divide, the growth-induced active stresses reorganize the topological defects within the colony, triggering active vortical flows within the colony. In experiments, the local nematic director and the position of the topological defects can be captured using time lapse imaging. The image data are analyzed using particle image velocimetry technique, and the emergent flows are visualized using a heat map that captures the flow magnitudes as shown in Fig.\ref{fig:stress}C. The relative strength of the active flow domains is indicated by the accompanying colour scale. The patchy flow landscape spatially correlates with the position of the topological defects, following the numerical predictions\cite{You:2018,Giomi:2014,Thampi:2014}, with the average patch size correlated with the mean separation of the topological defects. A closer look at the flow domains reveals the presence of counter-rotating vortical regions, shown in Figures \ref{fig:statistics}A \ref{fig:statistics}B, that emerge due to spontaneous disruption of the local orientational order. The regeneration and transformation of the topological defects allows sustained elastic and hydrodynamic interactions. In general, such flow fields can be broken down into radial and tangential components, as done using continuum modeling, shown in Fig. \ref{fig:statistics}A. Along the radial direction, the flow is predominantly expansive owing to the cell growth, whereas no net circulation is captured along the tangential direction\cite{You:2018}.

Recent experimental and theoretical studies in a range of living systems have shown that local order and topology can be crucial for biological functions at cellular and sub-cellular scales\cite{Hartmann:2019,Kawaguchi:2017,Saw:2017,Pawar:2014,Mathijssen:2018}.  The dynamics of phenotypic plasticity can be analysed in the framework of liquid crystallinity, where local anisotropy, order and topology underpins emergent mechanics at population, individual and sub-cellular scales. Despite the lasting evidence that liquid crystals are ubiquitous in, and intrinsic to, almost all biological structures\cite{Brown:1979}, their potential role in mediating phenotypic plasticity is largely unexplored, thus, leaving open a major gap in our efforts to understand the physics of life. Material topology, a salient attribute of liquid crystalline systems, goes beyond geometric shape, and can fundamentally impact the biophysics in microbial systems including mechanical pliability of colonies, transmission of mechanical stresses and transport of molecules and particles (e.g., bacteriophages), all of which are crucial determinants of a cell’s physiological state and fitness. A majority of current models of biologically active systems are based on particles possessing single, time-independent phenotypic traits: accounting for time-dependent phenotypes, observed frequently in natural and synthetic microbial systems, should allow development of new models that are not only richer in physics, but also a step closer to real microbial systems.

\section{\label{sec:feedback}Topological Feedback: Cross-talk between microbe and its micro-environment}

Microbes inhabit highly diverse ecosystems where periodic and stochastic variations in the environmental parameters determine cellular fitness, survival and succession. Light availability, ambient temperature, fluid flow and material compliance are typical abiotic factors that define the conditions of microbial environments (or the matrix). Unlike laboratory settings, where environmental factors are well-defined, controllable and tractable over time, natural parameters seldom represent steady state conditions. To bring out the functional role of active matter physics in microbial ecology, microorganisms need to be studied in relation to their environments, accounting for the spatial and temporal dynamics of the matrix attributes. From a biophysical standpoint, a vast majority of microbial matrices is composed of biopolymers, amphiphilic lipids, cytoskeletal and muscle proteins, collagens and proteoglycans, and liquid crystal phases (primarily, lyotropic or cholesteric phases)\cite{Sengupta:2015}. These building blocks for microbial matrices are inherently anisotropic, possessing long-range order with concomitant fluidity. The structured, out-of-equilibrium settings, combined with the inherent fluidity, render them akin to liquid crystal materials. More generally, variations in the environmental parameters can trigger topological transformations in the microbial matrix itself, thus initiating a topological feedback – an active cross-talk of underlying topological features of the microbial matter and the micro-environment. Conceptually, the framework for topological feedback was proposed recently by the author\cite{Gimoi:2017} in nematic micro-flows, where singularities across different fields (in this case, material and hydrodynamic fields) interacted in a coexisting setting. One could expect similar, yet much richer, singularity feedback at play in microbial ecosystems, wherein topological singularities - coupled across the microbial system and its surrounding matrix - lead to feedbacks that regulate microbial behaviour and physiology. 

Recent experiments along this line have been conducted with motile bacteria dispersed in liquid crystalline medium\cite{Peng:2016,Genkin:2017}. The results have demonstrated that microbial activity (motility in this case) couples with the topology of the local environment, ultimately biasing microbial migration. Depending upon the local topological characteristics, the cells were found to accumulate in the vicinity of $+1/2$ topological defects, and escape regions of topological $-1/2$ defect. The ability to regulate bacterial motion by imposing topological constraints in the surrounding environment offers a novel route to trap or transport natural and synthetic swimmers in anisotropic liquids. Consequently, patterns of topological defects could be further designed to tune the emergent order of surface-associated non-motile bacterial colonies or in populations of dense motile swimmers, to give rise to a novel class of active matter system. 

\begin{figure}[t]
\centering
\includegraphics[width=\columnwidth]{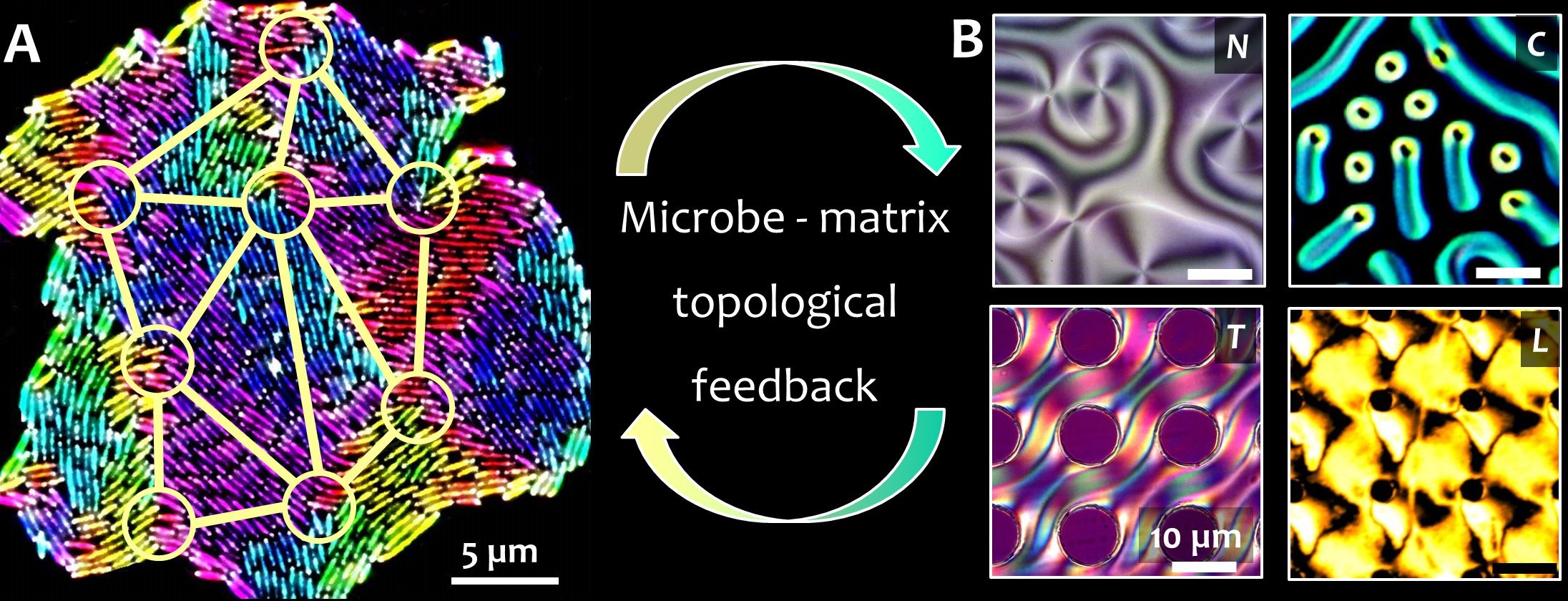}
\caption{\label{fig:coupling}{\bf Topological feedback in microbe-matrix ensemble.} (A) Topological defects can actively couple microbes and their surrounding matrix (micro-environment). For a growing bacterial colony, the anisotropy of the cell shape and the average growth rate determine the number of topological defects present at a given growth stage. (B) Alongside, the degree of order in the surrounding matrix, shown here for four distinct cases (clockwise, from top left) – nematic (N), cholesteric (C), and lyotropic liquid crystal materials, and a complex anisotropic substrate with topographical features (liquid crystal elastomer) – underpin the  strength of the topological coupling. Stronger the microbe-material coupling, more stable is the emergent topological feedback. This in combination with the growth induced mechanical stresses, introduces a novel biomechanical framework to analyze microbial physiology and behaviour, both of which can be tuned by the topology of the local micro-environment.}
\end{figure}

Beyond the exciting premise of engineering model active matter systems that can be tuned by matrix topology, or hydrodynamics or both, the coupling between microbial activity and matrix properties can have more fundamental implications. The dynamic feedback between the material and microbes can regulate the behaviour and physiology of microbes in a population or community, offering fitness benefits based on microbial phenotypes. As depicted in Fig.\ref{fig:coupling}, results are underway\cite{Sengupta:2020} that indicate organization of microbial topological defects in relation to the topological singularities in the surrounding matrix> The coupling interaction influences the spatial and temporal dynamics of the topological defects in the microbial colony. Since a growing colony of non-motile bacteria spontaneously forms a network of topological defects, the dynamics and organization of the defects is modified by the constraints posed by the surrounding matrix. Dynamic landscapes of topological defects in a microbial colony can cross-talk with the network of topological defects – imposed or spontaneously formed – in the surrounding matrix, influencing the active cell flow properties and molecular transport regimes. A number of liquid crystalline materials can be envisioned as potential test beds for studying microbe-matrix interactions (Fig.\ref{fig:coupling}B). When combined with micro-fabricated templates, we could arrive at a rich diversity of topological and topographical landscapes, over which microbial dynamics can be studied.

\section{\label{sec:conclusion}Conclusions and Prospects}

The advent of cross-disciplinary multi-scale experimental approaches has enabled simultaneous characterization of microbial behaviour, physiology and their habitats spanning multiple length and time scales. We are able to investigate and analyze living materials undergoing a major makeover – thanks to the physics of liquid crystals – that has propelled a growing exploration of topology-mediated physics in both fundamental studies and potential applications aimed at tailoring material attributes down to the molecular scale. Our ability to zoom into the micro-scale dynamics will help reveal how microbial environment – both structural and topological – shape the non-equilibrium dynamic, and over longer timescales, equilibrium microbial landscapes for single species or microbial consortia. Crucially, in a converse setting, we are on the verge of uncovering if (and how) micro-scale structural attributes in a given matrix – both topological and topographical – mediate microbial phenotypes, physiology and population fitness. A further boost in this direction could be provided by incorporating machine learning approaches to study microbe-material interfaces, including deep neural networks for feature recognition and tracking; or recurrent nets and random forests for analysis of time series. As highlighted in a recent Review\cite{Cichos:2020}, the promise and prospects of machine learning in active matter research is still in its infancy, however there is a growing interest in exploring this avenue, which has proved valuable in a number of other areas of research. By looking at the ensemble of microbes and their surrounding matrix, active topological feedback can be revealed, with potentially far-reaching implication in the fields of medical, food and environmental biotechnology. Furthermore, the activity of the microbes and the matrix can be varied - either selectively, or in unison through ecologically relevant biotic and abiotic factors. Taken together, the topological framework discussed here represents a rich and dynamic parameter space of material-microbe interactions, which is still awaiting exploration.

\section{\label{sec:acknow}Acknowledgements}

This work is supported by the ATTRACT Investigator Grant (grant A17/MS/11572821/MBRACE) of the Luxembourg National Research Fund. AS is grateful for the valuable discussions with L. Giomi, J. M. Yeomans, M. G. Mazza, W. C. K. Poon, M. M. Telo da Gama, F. Cichos, K. Kroy, C. Wagner, N. Araújo, J. Najafi, M. Ackermann, K. Drescher, J. Dunkel and members of the Physics of Living Matter Group, Luxembourg.

\end{document}